\documentstyle[12pt]{article}
\begin{document}
\newcommand{\be}{\begin{equation}}
\newcommand{\ee}{\end{equation}}
\newcommand{\vth}{\vspace{3mm}}
\newcommand{\osp}{osp(2|2)\oplus osp(2|2)}
\begin{flushright}
 RUP-94-17
\end{flushright}
\vspace{0.7cm}
\begin{center}
\Large{Topological Twist of $osp(2|2)\oplus osp(2|2)$\\
                 Conformal Super Algebra\\
                     in Two Dimensions}\\[0.7cm]
\large{Noriaki Ano
\footnote[1]{E-mail: nano@rikkyo.ac.jp}}\\
\small{\em Department of Physics, Rikkyo University}\\
\small{\em Nishi-Ikebukuro, Tokyo171, Japan}\\
\small{December, 1994}
\end{center}
\normalsize
\pagestyle{plain}
\vspace{1.0cm}
\begin{abstract}
A Lagrangian of the topological field theory is found in
the twisted $\osp$ conformal super algebra. The reduction on
a moduli space is then elaborated through the
vanishing Noether current.
\end{abstract}
\newpage
Since E. Witten has pioneered the topological field theory
(TFT)\cite{e.witten1},
many energetic researches have been
done\cite{d.birmingham&m.blau&m.rakowski&g.thompson1}
and then TFT
has been proved to be a real solid methodology
in the quantum field theory (QFT).
A few substantial problems associated with TFT still remain to be solved,
for example, about the topological twist, however.

There are two typical stand points for constructing TFT, i.e.
topological twisting and BRST gauge fixing.
Both approaches result in the so-called
moduli problem\cite{e.witten3}\cite{j.sonnenschein1}.
In either case, the remarkable characteristic is that the Lagrangian
is described as ${\cal L}=\{ Q,\enskip \star\}$, where $Q$ is the fermionic
operator of nilpotency. In terms of the ordinary
QFT words, ${\cal L}$ is just composed of the BRST gauge
fixing and the FP ghost terms, and the $Q$
corresponds to the BRST operator.
Because of the BRST-exact form of ${\cal L}$,
every correlation function is independent
of the coupling factor
as a consequence of which the leading contribution to the path-integral
is only the classical configuration of the fields, i.e. zero mode.
This zero mode configuration is associated with some moduli space.

In the BRST approach, the relation between the topological twist
and the moduli problem may be comparatively
clear owing to the intrinsic constructing
procedure where some moduli problem can be
settled as the gauge fixing condition.
In the topological twisting formalism,
the above relation is not much clear, on the contrary.
It seems that there has not been a
common recognition on what the topological twist is really doing.
The different twistings of the same model derive
the different moduli problems, respectively,
which are related through the mirror symmetry\cite{mirror}
as the explicit example of twisting in general.
In relation to the topological twisting mechanism,
the topological gauged WZW model\cite{h-l.hu1} is composed of two
different twisting procedures from the same bosonic model, not necessarily
twisting of the N=2 supersymmetry\cite{e.witten2}
of Kazama-Suzuki model\cite{y.kazama&h.suzuki1},
and in this case there surely exists
the mirror symmetry.

In this note, it is explicitly shown that the topological twist is related to
moduli problem through vanishing Noether current.
First of all, we perform topological
twisting of $\osp$ conformal super algebra.
The TFT Lagrangian is then constructed,
to be exact, found by using this twisted algebra
and the field configuration is investigated in the case
of zero-limit of the coupling factor on the path-integral.
This configuration is indeed a moduli space of flat connection,
and this fact originates from vanishing Noether current.\\

Let us start with explaining briefly
the reason why we use the conformal super group. The topological
twist usually means the mixing of the representation space of the automorphism
group of the supersymmetry with that of the symmetry group
with respect to the space-time, for example, spinor space. On the manifold,
the latter symmetry is local.
Consequently the former symmetry must also be local.
This situation is allowed only in the case of the
conformal supersymmetry.
This is the reason why we investigate this super group.
For the purpose of examining the
Yang-Mills theory on two dimensional manifold,
then, we are interested only in the symmetry group which
constructs a conformal supergravity theory\cite{j.mccabe&b.velikson1}.
The automorphism group of $Osp(2|2)\otimes Osp(2|2)$
in relation to $(2,2)$ supersymmetry is $SO(2)\otimes SO(2)$, and
$Osp(2|2)$ is required to be compact so that
its Cartan-Killing form is positive definite, while
the super Lie group $Osp(2|2)$ is generally not compact.

$\osp$ conformal super algebra on which the twisting operation will be made
to form a corresponding topological algebra
is then confined to the two dimensional
Lorentzian manifold with the local Lorentz metric
of the light cone coordinates:
$g^{z\bar{z}}=g^{\bar{z}z}=-2,\enskip g_{z\bar{z}}=g_{\bar{z}z}
=-1/2$ and $g^{zz}=g^{\bar{z}\bar{z}}=
g_{zz}=g_{\bar{z}\bar{z}}=0$.
This (2,2) super algebra contains two types of complex Weyl spinorial charges
$Q,\enskip \bar{Q};\enskip S,\enskip \bar{S}$,
where ``$-$'' means the Dirac conjugation
$\bar{Q}=Q^{\dagger}\gamma^0$ in which $\gamma^0=i\sigma^2$ and incidentally
$\gamma^1=\sigma^1,\enskip \gamma^5=-\gamma^0\gamma^1=\sigma^3$,
or equivalently
$\gamma^z=\gamma^0+\gamma^1,\enskip \gamma^{\bar{z}}=\gamma^0-\gamma^1$
in the light-cone coordinates.
These supercharges are two component spinors,
for example $Q=(Q_+,\enskip Q_-)^t$, where ``$+,-$'' mean spinor indices
describing ``left'' and  ``right'' moving
, respectively, with respect to the local Lorentz coordinates
$(z,\bar{z})$. These indices are raised and lowered
by a metric in spinor space
given by the charge conjugation matrix $C=\gamma^0$:
$\eta^{+-}=\eta_{-+}=-1,\quad \eta^{-+}=\eta_{+-}=1$ and
$\eta^{++}=\eta^{--}=\eta_{++}=\eta_{--}=0.$
We can leave out the conjugate
parts of the bra-ket bosonic relations
with respect to the complex supercharges of $\osp$ as follows:
\be
\begin{array}{c}
\mbox{$[S,P_a ]= \gamma_a Q,\hspace{1.0cm}
[S,D]=-\frac12 S,\hspace{1.0cm}[S,M]=-\frac12 \gamma_5 S,$}\\
\mbox{$[Q,K_a ]=-\gamma_a S,\hspace{1.0cm}
[Q,D]=\frac12 Q,\hspace{1.0cm}[Q,M]=-\frac12 \gamma_5 Q,$}\\
\mbox{$[S,A]=-i\frac14 \gamma_5 S,\hspace{1.5cm}[S,V]=-i\frac14 S,$}\\
\mbox{$[Q,A]=i\frac14 \gamma_5 Q,\hspace{1.5cm}[Q,V]=-i\frac14 Q.$}
\end{array}
\label{osp3}\ee
If we want to get these conjugate parts of Eqs.(\ref{osp3}),
we must pay attention to
the fact that the representation of
the body $o(2)\oplus sp(2)$
of $osp(2|2)$ are anti-Hermitian where the
anti-Hermitian character of the representation of $sp(2)$ actually
leads to the positivity of the Cartan-Killing form of $osp(2|2)$.

Ordinary (2,2) supersymmetry which is not conformal is direct sum
of (2,0) and (0,2)
and the corresponding part in $\osp$ reads
\be
\begin{array}{c}
\mbox{$\{Q_+,\bar{Q}_+\}=iP_z,\hspace{0.9cm}\{S_+,\bar{S}_+\}=-iK_z,$}\\
\mbox{$\{Q_-,\bar{Q}_-\}=iP_{\bar{z}},
\hspace{0.9cm}\{S_-,\bar{S}_-\}=-iK_{\bar{z}}.$}
\end{array}
\label{osp1}\ee
The super-extended conformal algebra on the other hand
has the central charges in the relations
between the super charges, and consequently
the decomposition mentioned above does not exist.
This mixing part of (2,0) and (0,2) in $\osp$ is
\be
\begin{array}{c}
\mbox{$\{Q_+,\bar{S}_-\}=i(M-D)+2(A-V),
\hspace{0.5cm}\{Q_-, \bar{S}_+ \}=i(M+D)+2(A+V),$}\\
\mbox{$\{\bar{Q}_+,S_-\}=i(M-D)-2(A-V),
\hspace{0.5cm}\{\bar{Q}_-,S_+ \}=i(M+D)-2(A+V).$}
\end{array}
\label{osp2}\ee
This property will play an important role in the forthcoming contexts.

The bosonic generators of $\osp$ are as follows:
$P_a,\enskip K_a,\enskip M,\enskip D,\enskip A$ and $V$ are
translation, conformal-translation, Lorentz, Weyl,
chiral $o(2)$ and internal $o(2)$, respectively.
The finite two-dimensional conformal algebra
composed of only these bosonic generators is
\be
\begin{array}{c}
\mbox{$[P_a,M]=\epsilon_{ab}P^b,
\hspace{1.5cm}[P_a,D]=P_a,\hspace{1.5cm}[K_a,D]=-K_a,$}\\
\mbox{$[K_a,M]=\epsilon_{ab}K^b,
\hspace{1.5cm}[K_a,P_b]=2(\epsilon_{ab}M-\delta_{ab}D),$}
\end{array}
\label{osp4}\ee
where the $\epsilon$ symbols are
$\epsilon^{z{\bar z}}=-\epsilon^{{\bar z}z}=-2,\enskip
\epsilon_{z{\bar z}}=-\epsilon_{{\bar z}z}=1/2$
and $\epsilon^{zz}=\epsilon^{\bar{z}\bar{z}}=
\epsilon_{zz}=\epsilon_{\bar{z}\bar{z}}=0$.\\

We are now in a position to perform
topological twisting of this algebra. Topological twist
is usually a kind of mixing which
results in identification of the representation space of
automorphism group of N=2 super symmetry with that of the local Lorentz group.
It is easy to perform
twisting of this algebra to get the topological algebra.
Most of all we have to do
is to replace $Q$, $\bar Q$, $S$, and $\bar S$ with
$Q^+$, $Q^-$, $S^+$, and $S^-$, respectively.
The indices ``$+,-$'' are raised and lowered with the same metric
as for the indices $\alpha ;\beta$ of $C_{\alpha \beta}$ and $Q_{\alpha}$.
That is, the complex Weyl spinors $\varphi_\alpha$,
$\bar{\varphi}_\alpha$ are substituted for
$\varphi_\alpha ^{\enskip +}$, $\varphi_\alpha^{\enskip -}$:
\be
\varphi_\alpha =\frac{i}{\sqrt{2}}\varphi_\alpha^{\enskip +},
\quad \bar{\varphi}_\alpha =
\frac{i}{\sqrt2} \varphi_\alpha ^{\enskip -}.
\label{topoto} \ee

The remaining manipulations are as follows. The fermionic charges
$Q^+=(Q_+^{\enskip +} ,Q_-^{\enskip +})^t$ have
become ( (0,0)-form , (0,1)-form ),
and $Q^-=(Q_+^{\enskip -} ,Q_-^{\enskip -})^t$
with ( (1,0)-form , (0,0)-form ),
{\it {idem}} $S^{\pm}$.  Then we have to
modify the definitions of local Lorentz $M$
and Weyl $D$ generators so that the four (0,0)-form fermionic generators
of super symmetry have no charge with respect to
these two bosonic generators. We have put the representation
space accompanied with the automorphism group $O(2)\otimes O(2)$ upon
the space of spinor. The modified $M$, $D$ generators must be direct sums
with $o(2)\oplus o(2)$
generators $V$ and  $A$, respectively. The solution to this
constraint resolves into
\be
\begin{array}{cc}
\tilde{M} =M+2iV, & \tilde{D} =D+2iA.
\end{array}
\label{c}
\ee
These modified generators then satisfy the following relations:
\be
\begin{array}{cc}
[ Q_{\pm}^{\enskip \pm}, \tilde{M} ]=0,
& [ S_{\pm}^{\enskip \pm}, \tilde{D} ]=0.
\end{array}
\label{d}
\ee

There appear some problems about the closure of the modified algebra, however.
The generators $A$ and $ V $ have been put upon $D$ and $M$, respectively
and the modified algebra which contains $\tilde{M}$ and $\tilde{D}$
must not contain $A$ and $ V $.
In fact, the modified algebra contains subtle relations:
\[
\{Q_-^{\enskip +}, S_+^{\enskip -}\}=i(\tilde{M}+\tilde{D})-4i(A+V),
\]
\be
\{Q_+^{\enskip -}, S_-^{\enskip +}\}=i(\tilde{M}-\tilde{D})+4i(A-V).
\label{e}
\ee

We can neglect the constraints(\ref{e}) as in the followings.
Here it is necessary to omit
another generators with regard to Eqs.(\ref{e}),
if this modified algebra still obeys the closure
property for the generators of the gauge symmetry.
In this point of view, the four fermionic generators
$Q_+^{\enskip -}$, $Q_-^{\enskip +}$, $S_+^{\enskip -}$ and $S_-^{\enskip +}$
do not induce the gauge transformations generated by both
$i(\tilde{M}+\tilde{D})-4i(A+V)$ and $i(\tilde{M}-\tilde{D})+4i(A-V)$.
To do so, there are two alternatives, that is,
the case in which the holomorphic vector charges
$Q_+^{\enskip -}, S_+^{\enskip -}, P_z,$ and $K_z$
vanish, or the case in which the anti-holomorphic charges
$Q_-^{\enskip +}, S_-^{\enskip +}, P_{\bar{z}},$ and $K_{\bar{z}}$
vanish. The second case is adapted here without any compensation procedure,
that is, all gauge fields and parameters of these four generators are
assured to vanish.

This twisting procedure is explained in terms of the gauge fields of the
corresponding symmetry $\osp$. Let us introduce the gauge
field ${\bf a}$ which is Lie super algebra-valued
1-form of $\osp$ in the form
\begin{eqnarray}
{\bf a}_\mu = e_\mu ^a P_a & +&  f_\mu ^a K_a  +
\omega_\mu M + b_\mu D \nonumber \\
  & + & a_\mu A + v_\mu V +\bar{\psi}_\mu Q + \bar{Q} \psi_\mu +
\bar{\phi}_{\mu} S + \bar{S} \phi_\mu,
\end{eqnarray}
as well as transformation parameter ${\bf {\tau}}$ defined by
\begin{eqnarray}
{\bf {\tau}} = \xi_P ^a P_a & + & \xi_K ^a K_a  +
\lambda_l M + \lambda_d D \nonumber \\
& + & \theta_a A + \theta_v V +
\bar{\varepsilon} Q + \bar{Q} \varepsilon
+ \bar{\kappa} S + \bar{S} \kappa.
\label{para}\end{eqnarray}
Using these gauge fields and parameters, the above mentioned
topological twist and additional manipulations can be described as follows:
Eqs.(\ref{c}) mean
\be
v_\mu=2i\omega_\mu, \quad a_\mu = 2ib_\mu,
\label{modcon} \ee
and neglect of the generators $Q_-^{\enskip +}$,
$S_-^{\enskip +}$, $P_{\bar{z}}$
and $K_{\bar{z}}$ means
\begin{eqnarray} \vth
\phi_{\mu +}^{\quad -} = 0 = \psi_{\mu +}^{\quad -},&\quad &
\kappa_+^{\quad -} = 0 = \varepsilon_+^{\quad -},\nonumber \\ \vth
e_\mu ^{\bar{z}} = 0 = f_\mu ^{\bar{z}},&\quad &
\xi_P ^{\bar{z}} = 0 = \xi_K ^{\bar{z}}.
\label{figcon} \end{eqnarray}

Under these conditions we are led to
\be
\delta \phi_{\mu +}^{\quad +} \sim \delta \phi_{\mu -}^{\quad -}, \quad
\delta \psi_{\mu +}^{\quad +} \sim \delta \psi_{\mu -}^{\quad -}.
\label{equiv} \ee
Accordingly we have the following identifications:
\[ \psi_{\mu +}^{\quad +} = -\psi_{\mu -}^{\quad -} \equiv -\psi,\quad
\phi_{\mu +}^{\quad +} = -\phi_{\mu -}^{\quad -} \equiv -\phi,\]
\be
\varepsilon_+^{\enskip +} = -\varepsilon_-^{\enskip -}
\equiv -\varepsilon,\quad
\kappa_+^{\enskip +} = -\kappa_-^{\enskip -} \equiv -\kappa,
\label{id} \ee
which read without loss of generality
\be
Q \equiv Q_+^{\enskip +} + Q_-^{\enskip -},\quad S
\equiv S_+^{\enskip +} + S_-^{\enskip -}.
\label{cha} \ee
Taking all these additional conditions with respect to topological twisting of
the original $\osp$ into account,
we get the gauge connection ${\bf a}$:
\begin{eqnarray}
{\bf a}_{\mu}=e_{\mu}^z P_z + f_{\mu}^z K_z &+&
\omega_{\mu} \tilde{M} + b_{\mu} \tilde{D} \nonumber\\
 &-& \frac12 (Q_+^{\enskip -}\psi_{\mu -}^{\quad +} +
S_+^{\enskip -}\phi_{\mu -}^{\quad +}
+ \psi_{\mu} Q + \phi_{\mu} S),\label{fineqc}\end{eqnarray}
and transformation parameter ${\bf {\tau}}$:
\begin{eqnarray}
{\bf {\tau}}= \xi_P^z P_z + \xi_K^z K_z &+&
\lambda_l \tilde{M} + \lambda_d \tilde{D}\nonumber \\
&-& \frac12 (Q_+^{\enskip -}\varepsilon_-^{\enskip +}
+ S_+^{\enskip -}\kappa_-^{\enskip +}
+ \varepsilon Q +\kappa S),\label{fineq} \end{eqnarray}
respectively.

After all, the generators in
Eqs.(\ref{fineqc})(\ref{fineq}) obey the following relations:
\be
\begin{array}{lll}
\mbox{$[S,P_z]= Q_+^{\enskip -},$}&
\mbox{$[Q_+^{\enskip -},\tilde{D}]=Q_+^{\enskip -},$}&
\mbox{$[Q_+^{\enskip -},\tilde{M}]=-Q_+^{\enskip -},$}\\
\mbox{$[Q,K_z]=-S_+^{\enskip -}$}&
\mbox{$[S_+^{\enskip -},\tilde{D}]=-S_+^{\enskip -},$}&
\mbox{$[S_+^{\enskip -},\tilde{M}]=-S_+^{\enskip -},$}\\
\mbox{$\{ Q,Q_+^{\enskip -} \} =-2iP_z,$}&
\mbox{$\{S,S_+^{\enskip -} \} =2iK_z,$}&
\mbox{$\{ Q,S \} = -4i\tilde{M},$}\\
\mbox{$[P_z, \tilde{M}]=-P_z,$}&\mbox{$[P_z, \tilde{D}]=P_z,$}&\quad\\
\mbox{$[K_z,\tilde{M}]=-K_z,$}&\mbox{$[K_z,\tilde{D}]=-K_z,$}&\quad\\
\end{array}
\label{topoal}\ee
and the gauge connections satisfy the following transformation rules:
\be
\begin{array}{l} \vth
\mbox{$\delta\psi_\mu=\partial_\mu\varepsilon$},\\ \vth
\mbox{$\delta \phi_\mu = \partial_\mu \kappa $},\\ \vth
\mbox{$\delta \psi_{\mu -}^{\quad +} = \xi_P^z \phi_\mu +
( \lambda_l - \lambda_d )
\psi_{\mu -}^{\quad +} - e_\mu^z \kappa
+ {\cal D}_{\mu}\varepsilon_-^{\enskip +} $},\\ \vth
\mbox{$\delta \phi_{\mu -}^{\quad +} = -\xi_K^z
\psi_\mu + ( \lambda_l + \lambda_d)
\phi_{\mu -}^{\quad +} + f_\mu^z \varepsilon
+ {\cal D}{\mu}\kappa_-^{\enskip +} $},\\ \vth
\mbox{$\delta e_\mu^z ={\cal D_\mu}\xi_P^z +
(\lambda_l - \lambda_d) e_\mu^z
+\frac{i}2 (\varepsilon \psi_-^{\enskip +}-
\psi \varepsilon_-^{\enskip +}) $},\\ \vth
\mbox{$\delta f_\mu^z={\cal D}_\mu \xi_K^z +
(\lambda_l + \lambda_d) f_\mu^z -\frac{i}2
(\kappa \phi_-^{\enskip +} - \phi \kappa_-^{\enskip +}) $},\\ \vth
\mbox{$\delta \omega_\mu = {\cal D}{\mu}\lambda_l
+\frac{i}2 (\kappa \psi_{\mu} + \varepsilon \phi_{\mu}),$}\\ \vth
\mbox{$\delta b_\mu = {\cal D}\lambda_d,$}
\end{array}
\label{gaugeal} \ee
where
\be
\begin{array}{c}
\mbox{$ {\cal D}\varepsilon_-^{\enskip +}=
(\partial_\mu + b_\mu -\omega_\mu)\varepsilon_-^{\enskip +},\quad
{\cal D}\kappa_-^{\enskip +}=
(\partial_\mu - \omega_\mu - b_\mu)\kappa_-^{\enskip +},$}\\
\mbox{$ {\cal D}_\mu \xi_P^z =
(\partial_\mu - \omega_\mu + b_\mu)\xi_P^z,\quad
{\cal D}_\mu \xi_K^z = (\partial_\mu - \omega_\mu - b_\mu)\xi_K^z, $}\\
\mbox{$ {\cal D}_{\mu}\lambda_l=\partial_{\mu}\lambda_l,\qquad
{\cal D}_{\mu}\lambda_d=\partial_{\mu}\lambda_d.$}
\end{array}
\ee
The field strengths in relation to the discarded $Q_-^{\enskip +}$,
$S_-^{\enskip +}$, $P_{\bar{z}}$ and $K_{\bar{z}}$ all vanish as expected.
The resultant algebra (\ref{topoal}) can be
referred to topological algebra\cite{j.labastida&p.llatas1}.

$o(2)\oplus o(2)$ generators,
$A$ and $V$, still remain as global internal
symmetry whose charge is the so-called ghost-number,
the generators of which are defined by $G\equiv 2i(A-V)$,
$\tilde{G}\equiv 2i(A+V)$.
Here $G$ and $\tilde{G}$ satisfy the following relations:
\be
\begin{array}{cc}
\mbox{$[G,Q_+^{\enskip +}]=Q_+^{\enskip +},$} &
\mbox{$[G,Q_+^{\enskip -}]=-Q_+^{\enskip -},$} \\
\mbox{$[\tilde{G},Q_-^{\enskip -}]=Q_-^{\enskip -},$} &
\mbox{$[\tilde{G},Q_-^{\enskip +}]=Q_-^{\enskip +},$}\\
\mbox{$[G,S_+^{\enskip +}]=-S_+^{\enskip +},$} &
\mbox{$[G,S_-^{\enskip +}]=S_-^{\enskip +},$} \\
\mbox{$[\tilde{G},S_-^{\enskip -}]=-S_-^{\enskip -},$} &
\mbox{$[\tilde{G},S_+^{\enskip -}]=S_+^{\enskip -}.$}
\end{array}
\label{i}
\ee
As a consequence of Eqs.(\ref{i}), indeed,
it is natural to regard these generators $G$, $\tilde{G}$
as the ghost number operators.
$Q_{\pm}^{\enskip \pm}$ and $S_{\pm}^{\enskip \mp}$ increase the ghost
number by one unit, while $Q_{\pm}^{\enskip \mp}$
and $S_{\pm}^{\enskip \pm}$ decrease it by the
same quantity.
This assignment is consistent with the
relations(\ref{topoal})(\ref{gaugeal}).\\

Hereafter, our principal concern is now reduced to building up
the TFT Lagrangian and quantizing it.
Let us give a brief sketch of
TFT\cite{d.birmingham&m.blau&m.rakowski&g.thompson1}.
The theory has some fermionic operator
$Q$ with its nilpotency and the Lagrangian $\cal L$ is
$Q$-exact: ${\cal L}=\{ Q, \star \}$.
Moreover the TFT Lagrangian is allowed
to have the gauge symmetries, in which the cohomological nature of TFT
turns out to the equivariant cohomology:
$ Q^2=\tau_{\phi}$, where $\tau_{\phi}$
means the gauge transformation with the parameter $\phi$.
The freedom of this gauge symmetry can be fixed by using
BRST method as usual before the quantization of this system.
But in TFT this freedom may be automatically fixed
after the quantization, i.e. the reduction on some moduli space.
Because ${\cal L}$ is $Q$-exact, the correlation functions
are independent of a coupling factor.
The zero limit of the coupling factor then
induces the leading contribution of the path-integral with ${\cal L} =0$.
This configuration of the fields is associated with a moduli space.
As a consequence, TFT makes a local field theory
in the sense of a finite integration
on the moduli space after quantization.
Parenthetically it is necessary to mention that
the behavior of the path-integral measure is
crucial for the case of the full construction
of TFT\cite{e.witten2}, e.g. in studies of the mirror
symmetry\cite{e.witten4}\cite{p.aspinwall&d.morrison1}.\\

Let us turn attention to building up TFT Lagrangian.
We can construct
${\cal L}$ by using the anti-commutation relations in (\ref{topoal}):
\be
{\cal L} \equiv \{ Q , S \},
\label{j}
\ee
because of
\be
\{Q,S\}=\int_{\partial M^2}J_{ \{Q,S \} }^0=\int_{M^2}dJ_{ \{Q,S \} }^0,
\ee\\
where $Q=Q_+^{\enskip +} + Q_-^{\enskip -},
\enskip S=S_+^{\enskip +} + S_-^{\enskip -}$,
and $Q^2 =0=S^2$.
It is of interest to mention that ${\cal L}$ can be defined only in the
two dimensional surface with boundary,
has no ghost number and is invariant under the gauge symmetries
generated by $\tilde{M}$, $\tilde{D}$, $Q$, and $S$.
We can quantize formally this system of
Lagrangian(\ref{j}) by the path integral.
It is well known that the path-integral on the manifold
with boundary describes some states on the boundary,
which are also topological invariants\cite{e.witten1},
and consequently we can make the formulation
on the manifold without boundary by means of the inner
product of the path-integrals as ``in'' and
``out'' states. Hereafter we then suppose that the manifold
of the theory would be without boundary\cite{m.blau&g.thompson1}.

As mentioned above, the configuration of the system results in a
corresponding moduli space after quantization.
Therefore we obtain
\be
\ { \cal L } =  \{ Q,S \} \nonumber  =  -4i{\tilde{M}} = 0.
\label{k}
\ee
Suppose there exists a system $\Gamma$ of the gauge fields alone
and let the Noether current of the symmetry
generated by ${\tilde{M}}$ be $J_0^{\tilde{M}}$,
the condition (\ref{k}) turns out to $J_0^{\tilde{M}}$=0
which is the result of the path integral.
Consequently we define a sub-configuration
$\Gamma_s$ which satisfies $J_0^{\bar{M}}=0$.
The constraint $J_0^{\tilde{M}}$=0 provides us
with the following reduction of the configuration:
\be
\Gamma \quad \Longrightarrow \quad \Gamma_s.
\label{l}
\ee

Let us confine ourselves to investigation of the physical meaning
of the reduction mentioned above just through the
Noether current which is composed of the connections of
original $Osp(2|2)\otimes Osp(2|2) (\equiv{\cal G})$.
Let us start with considering the Yang-Mills action on
2-dimensional manifold without boundary:
\be
{\cal L}_{YM_2}=\int_{M^2} |R_{\cal G}|^2 \ast 1,
\label{ym} \ee
where $\ast$ is Hodge star operator and
$R_{\cal G}$ is a field strength 2-form :$R_{\cal G}=R^A B I_{AB}$
in which $I_{AB}$ is
the Cartan-Killing form on the Lie algebra of ${\cal G}$:
$I_{AB}=f_{AC}^{\quad D}f_{BD}^{\quad C}$.
It is a matter of course that Eq.(\ref{ym}) is invariant with respect to
the gauge symmetry ${\cal G}$ and has no general coordinate invariance.
Here the norm $|R_{\cal G}|^2$ has been
obtained by using the metric on $M^2$ and the Killing form $I_{AB}$.
It is of interest to argue that the integrand
of Eq.(\ref{ym}) can be rewritten as
$R^A \wedge \ast R^B I_{AB}$, together
with the volume form of the metric  $\ast 1$.
The time component of the Noether current
in association with the symmetry generated by
$\tilde{M}=M+2iV$ then turns out to be
\be
J_{\tilde{M}}^0 = \frac{\partial {\cal L}_{YM_2}}
{\partial(\partial_0{\bf a}_{\mu}^A)}
G_{\tilde{M}}^A({\bf a})_{\mu}=R^BI_{AB}G_{\tilde{M}}^A({\bf a})_{\mu},
\label{current} \ee
where $G_{\tilde{M}}^A({\bf a})_{\mu}$ are defined by
\be
\delta_{\tilde{M}}{\bf a}_{\mu}^A=\lambda_{\tilde{M}}
G_{\tilde{M}}^A({\bf a})_{\mu}.
\label{gg}\ee
${\bf a}^A$ are general form of the connections ${\bf a}_{\mu}^A$
and $R^A\equiv R_{01}^A$.
Parenthetically as be seen from Eq.(\ref{current}),
$\tilde{M}$ does not mean the topological twist in the situation.

We can now add the optional field strength components to the original current
owing to the ambiguity of the Noether current.
We are free to choose the additional terms:
\be
\partial_{\alpha}(\theta^AR^{A \alpha\beta}),
\label{add}\ee
where $\theta^A(x)$ 0-form are arbitrary functions which supplement
the characteristics of $J_{\tilde{M}}^0$
with respect to the paired field strengths $R^A$.
For the purpose of determining the compensating factors $\theta^A(x)$ 0-form,
we refer to Eq.(\ref{current}) in which
$G_{\tilde{M}}^A({\bf a})_{\mu}$ are composed of
the gauge transformations $\delta_{/{\bf \tau}}{\bf a}^A$
1-form as in Eqs.(\ref{gg}), and surely
correspond to $\theta^A(x)$ 0-form.
Here $\theta^A(x)$ 0-form must be constructed through the
reduction procedure of $\delta{\bf a}^A$ 1-form.
That is, we need some map $X$:
\be
X:\delta{\bf a}^A \mbox{...1-form}\quad\mapsto\quad\theta^A \mbox{...0-form}.
\ee
This map $X$ can indeed be chosen as
\be
X\equiv{\cal D}^{\ast},
\ee
where ${\cal D}^{\ast}$ is the adjoint exterior derivative operator:
\be
{\cal D}^{\ast}:\Omega^r(M)\rightarrow\Omega^{r-1}(M),
\ee
with ${\cal D}^{\ast}=(-1)^{2r}\ast
{\cal D}\ast$ on the two dimensional Lorantzian manifold
without boundary. We then obtain $\theta^A(x)$ 0-form as follows:
\be
\theta^A\equiv{\cal D}^{\ast}\delta{\bf a}^A.
\ee
Accordingly Eqs.(\ref{add}) are reduced to
\be
\partial_{\beta}(\theta^A R^{A\beta\alpha})=
(\partial_{\beta}\theta^A-(-)^{|B||A|}f_{BA}^{\quad C}
\theta^C{\bf a}_{\beta}^B)R^{A\beta\alpha}
+\theta^A{\cal D}_{\beta}R^{A\beta\alpha},
\label{addpa}\ee
where
\be
{\cal D}_{\beta}R^{A\beta\alpha}=\partial_{\beta}R^{A\beta\alpha}+
(-)^{|B||C|}f_{BC}^{\quad A}{\bf a}_{\beta}^B R^{C\beta\alpha}.
\ee
We next add Eqs.(\ref{addpa})
to $J_{\tilde{M}}^0$ which leads to
\be
J_{\tilde{M}}^0=\sum_A^{all}(\Pi_{\mu=1}^AR^A+\theta^A{\cal D}_{\mu=1}R^A),
\label{jj}\ee
where
\be
\Pi_{\mu=1}^A\equiv G_{\bar{M}}^A({\bf a})_1+
\partial_1\theta^A-(-)^{|B||A|}f_{BA}^{\quad C}\theta^C{\bf a}_1^B.
\ee

Our principal task is now to make the topological twist
of the current (\ref{jj}) and set it upon the configuration $\Gamma_s$.
The zero field strengths on $\Gamma_s$ are removed through replacement
of $A$ by $A^{\ast}$ which is defined by
\be
A^{\ast}=P_z,K_z,\tilde{M},\tilde{D},Q,Q_+^{\enskip -},S,S_+^{\enskip -}.
\label{info}
\ee
We then obtain the informations for $J_{\tilde{M}}^0=0$ as follows:
\be
R^{A^{\ast}}=0,\qquad\theta^{A^{\ast}}=0.
\label{condition}\ee
Clearly this solution(\ref{condition}) is not unique
in the mathematical view point,
but seems natural because of the independence of the
specific space-time coordinate index:$\mu=1$.

$R^{A^{\ast}}=0$ and $\theta^{A^{\ast}}=0$ obtained above play the
roles of the constraints for the configuration ${\Gamma}_s$
which eventually lead to some moduli space.
Number of the equivariant constraints
$R^{A^{\ast}}=0$ is equal to that of the
fermionic connections with ghost number
$\psi_{\mu}(-1)$,\enskip $\psi_{\mu -}^{\quad +}(1)$,\enskip
$\phi_{\mu}(1)$ and $\phi_{\mu -}^{\quad +}(-1)$.
Therefore these constraints can be regarded as the fixing conditions
of the so-called topological symmetry whose degrees of freedom
is equal to number of these fermionic connections,
i.e. the so-called topological ghosts.

Let us next explain the physical meaning of
the conditions $\theta^{A^{\ast}}=0$.
The tangent of the connection space ${\cal A}$
can be decomposed\cite{o.babelon&c.m.viallet1} as follows:
\be
T_{\bf a}{\cal A}=Im{\cal D}\oplus ker{\cal D^{\ast}},
\ee
where $Im{\cal D}$ 1-form is the tangent in the gauge
direction, while $Ker{\cal D}^{\ast}$ 1-form means the
component orthogonal to the gauge orbit and $0=\theta^{A^{\ast}}=
{\cal D}^{\ast}\delta{\bf a}^{A^{\ast}}$ is a
natural gauge condition in which $\delta{\bf a}^{A^{\ast}}$ is an
infinitesimal variation of the connection ${\bf a}$.
The constraints $\theta^{A^{\ast}}$, the number of which is equal
to that of the generators of the gauge symmetry,
is then regarded as the gauge fixing conditions.

We can therefore claim that
all the constraints $R^{A^{\ast}}=0$,\enskip$\theta^{A^{\ast}}=0$ which
originate from $J_{\tilde{M}}^0=0$ indeed lead to a moduli space
of flat connection:
\be
{\cal M}_{flat}=\{R^{A^{\ast}}=0\}/{\cal G^{\ast}}.
\label{modu}\ee
The BRST gauge fixing is necessary,
by way of parenthesis,
for the detailed investigation of the observables,
correlation functions and
their geometrical meaning in TFT\cite{h.kanno1}\cite{l.baulieu&i.singer1}.
Incidentally let us describe another representation
for the conditions (\ref{condition}).
If the infinitesimal variation of the connections $\delta{\bf a}^{A^{\ast}}$
are on $\Gamma_s$, the variations of $R^{A^{\ast}}$
under $\delta{\bf a}^{A^{\ast}}$ must also vanish.
Linearized representation\cite{m.f.atiyha&r.bott1}\cite{n.j.hitchin1}
of the flat connection equations yields
\be
\begin{array}{l}
\mbox{$0=\ast\delta R=\ast{\cal D}\delta {\bf a}=
\ast{\cal D}\ast\ast\delta{\bf a}={\cal D}^{\ast}\ast\delta{\bf a},$}\\
\mbox{${\cal D}^{\ast}\delta{\bf a}=0.$}
\end{array}
\ee
If $\delta{\bf a}$ are the arbitrary variations on ${\cal M}_{flat}$,
its Hodge dual $\ast\delta{\bf a}$, which are
still 1-forms only in the two-dimension, are also on ${\cal M}_{flat}$.

Let us shed some light upon the principal result (\ref{modu}).
In the ordinary N=2 conformal supergravity, all
curvatures must vanish in full consonance with the general coordinate
transformations as the gauge symmetry generated by the conformal
super group\cite{supergravityconstraint}.
As a consequence, there exists no kinetic term, i.e.
no dynamics of the connection fields in the ordinary theory.
In the present case, on the contrary,
zero curvatures play the roles of the conditions
which lead to the configuration of the fields.
Accordingly the general coordinate transformations
$\delta_{gc}(\xi)$ are
induced by these conditions.
$\delta_{gc}(\xi)$ expressed as
\be
\delta_{gc}(\xi){\bf a}_{\mu}^A=\sum_B\delta_B(\xi^{\nu}{\bf a}_{\nu}^B)
{\bf a}_{\mu}^A+\xi^{\nu}R_{\nu\mu}^A.
\label{geneco}\ee
The topological twist of Eqs.(\ref{geneco}) then yields
\be
\delta_{gc}(\xi){\bf a}_{\mu}^{A^{\ast}}=\sum_{B^{\ast}}\delta_{B^{\ast}}
(\xi^{\nu}{\bf a}_{\nu}^{B^{\ast}}){\bf a}_{\mu}^{A^{\ast}},
\ee
with the additional use of Eqs.(\ref{condition}).
$\delta_{gc}(\xi)$ have been fixed in accordance with
the thoroughly fixed gauge symmetry (\ref{modu}).
It is then possible to argue that
the configuration ${\cal M}_{flat}$ is a quotient
not only in the sense of the gauge symmetry,
but also in the sense of the diffeomorphism:
\be
\sim / {\cal G}^{\ast}\quad \supset \quad \sim / \mbox{Diff}_0.
\ee\\

Let us make concluding remarks as follows:
{\it First}, making use of
the ambiguity of the Noether current,
we are led to the relation (\ref{jj}) between the vanishing Noether
current $J_{\tilde{M}}^0=0$ and the flat connection
conditions $R^{A^{\ast}}=0, \theta^{A^{\ast}}=0$.
In the conventional QFT, this ambiguity mentioned above
plays an important role in association with
avoiding the anomaly, while its physical role is
not clarified in the case of the classical correspondent.
In the present TFT, on the contrary,
the classical theory has been obtained as
the limiting case of the path-integral,
and consequently the ambiguity argued above leads to
the corresponding moduli problem.
{\it Secondly}, we are led to claim that the
vanishing Noether current $J_{\tilde{M}}^0=0$ induces the field
configuration ${\cal M}_{flat}$ after
the path-integral when the Lagrangian is constructed from the
twisted algebra (\ref{topoal}) and that
the vanishing Noether current $J_{\tilde{M}}^0=0$
plays the role equivalent to the moduli problems.
On the other hand, since physical aspects of CFT
are dominantly described by using the canonical formulation
of currents, the vanishing Noether current $J_{\tilde{M}}^0=0$
may shed some light upon establishing the deeper insight
into CFT in general.
The characteristic feature of the current
associated with $\tilde{M}$ will indeed be
clarified in the matter systems.
\vspace{1.0cm}\\

The author would like to thank Professor H.Fujisaki
for enlightening discussions and careful reading of the manuscript.
He is much grateful to Dr.T.Suzuki
for fruitful discussions and helpful comments.
He is also indebted to Mr.T.Nonaka for useful discussions.\\



\begin{thebibliography}{99}

\bibitem{e.witten1} E. Witten, Commun. Math. Phys. {\bf 117}(1988)353.
\bibitem{d.birmingham&m.blau&m.rakowski&g.thompson1}See for example,
D. Birmingham, M. Blau, M. Rakowski and G. Thompson,
Phys. Rep. 209(1991)129, and references therein.
\bibitem{e.witten3} E. Witten, Int. J. Mod. Phys. {\bf A6}(1991)2775.
\bibitem{j.sonnenschein1} J. Sonnenschein, Phys. Rev. {\bf D42}(1990)2080.
\bibitem{mirror} P. Candelas, M. Lynker
and R. Schimmrigk, Nucl. Phys. {\bf B341}(1990)383.\\
M. Lynker and R. Schimmrigk, Phys. Lett. {\bf B249}(1990)237.\\
B. R. Greene and R. Plesser, Nucl. Phys. {\bf B338}(1990)15.
\bibitem{h-l.hu1} H-L. Hu, Phys. Rev. {\bf D46}(1992)1761.
\bibitem{e.witten2} E. Witten, Nucl.Phys. {\bf B371}(1992)191.
\bibitem{y.kazama&h.suzuki1} Y. Kazama and
H. Suzuki, Nucl. Phys. {\bf B321}(1989)232.
\bibitem{j.mccabe&b.velikson1} J. MacCabe and
B. Velikson, Phys. Rev. {\bf D40}(1989)400.
\bibitem{j.labastida&p.llatas1} J. M. F. Labastida and
P. M. Llatas, Nucl. Phys. {\bf B379}(1992)220.
\bibitem{e.witten4} E. Witten, {\it Mirror Manifolds and Topological
Field Theory}, in: Essays on Mirror Manifolds (International Press, 1992),
hep-th/9112056.
\bibitem{p.aspinwall&d.morrison1} P. Aspinwall and
D. Morrison, Commun. Math. Phys. {\bf 151}(1993)245.
\bibitem{m.blau&g.thompson1}See for example,
G. Thompson, {\it 1992 Trieste Lectures on
Topological Gauge Theory And Yang-Mills Theory},
in: Proc.High Energy Physics and Cosmology (Trieste, 1992), hep-th/9305120,
and references therein.
\bibitem{o.babelon&c.m.viallet1} O. Babelon and
C. M. Viallet, Commun. Math. Phys. {\bf 81}(1981)515.
\bibitem{h.kanno1} H. Kanno, Z. Phys. {\bf C43}(1989)477.
\bibitem{l.baulieu&i.singer1} L. Baulieu and
I. Singer, Nucl. Phys. B(Proc. Suppl.){\bf 5B}(1988)12.
\bibitem{m.f.atiyha&r.bott1} M. Atiyha and
R. Bott, Philos. Trans. R. Soc. London {\bf A308}(1982)523.
\bibitem{n.j.hitchin1} N. J. Hitchin, Proc. London Math. Soc. {\bf 55}(1987)59.
\bibitem{supergravityconstraint} P. van Nieuwenhoizen,
Int. J. Mod. Phys. {\bf A1}(1986)155. \\
M.Hayashi, S. Nojiri and S. Uehara, Z. Phys. {\bf C31}(1986)561.
\end{thebibliography}
\end{document}